\documentclass[aps,pre,showpacs,manuscript,tightenlines]{revtex4}
\usepackage[centertags]{amsmath}
\usepackage{amssymb}
\usepackage{amsthm}
\usepackage{graphicx,subfigure}
\usepackage{epsfig}
\usepackage{multirow}
\usepackage{color}

\begin{document}
\title{The Fokker Planck equation for the non-Markovian harmonic oscillator across a magnetic field}

\author{Joydip Das and Bidhan Chandra Bag
{\footnote{Author for correspondence,
e-mail:bidhanchandra.bag@visva-bharati.ac.in}}}

\affiliation{Department of Chemistry, Visva-Bharati, Santiniketan 731 235, India}

\begin{abstract}
In a recent paper Das {\it et al.} [J. Chem. Phys. {\bf 147}, 164102 (2017)] proposed the Fokker-Planck equation (FPE) for the Brownian harmonic oscillator in the presence of magnetic field and the non-Markovian thermal bath, respectively. This system has been studied very recently by
Hidalgo-Gonzalez and  Jim\'enez-Aquino [Phys. Rev. E {\bf{100}}, 062102 (2019)] and the Fokker-Planck equation was derived  using the characteristic function. It includes a few extra terms in the FPE and the authors conclude that their method is accurate compared to the calculation by Das {\it et al.}. Then we reexamine our calculation and which is present in this comment. The revised calculation shows that both the methods give the same result.   
\end{abstract}


\maketitle

In a recent paper \cite{das}, we derived the Fokker-Planck equations using an alternative method for the non-Markovian dynamics for a free particle and  the harmonic oscillator, respectively. Then we extend this method for the non-Markovian dynamics in the presence of a magnetic field. Very recently,
the FPE equation for a non-Markovian harmonic  oscillator across a magnetic field has been derived by the characteristic function in Ref.\cite{preha}. Here it has been shown that a few extra terms appear in the FPE compared to Ref.\cite{das}. Then to reexamine our calculation we started with the relevant Langevin equations of motion \cite{das,preha},

\begin{equation} \label{leq1}
\dot{u}_x= -\omega^2 x-\int_0 ^t\gamma(t-\tau) u_x (\tau)d\tau+ \Omega u_y + f_x (t)   
\end{equation}
\noindent
and
\begin{equation} \label{eqq}
\dot{u}_y = -\omega^2 y-\int_0 ^t\gamma(t-\tau) u_y (\tau)d\tau- \Omega u_x + f_y (t)   \; \; \;,  
\end{equation} 
\noindent
where $\omega$ is the frequency of the Harmonic Oscillator and $\Omega$ corresponds to the cyclotron frequency. The random forces, $f_x$ and $f_y$ are independent  Gaussian noises  and they are related with the frictional memory kernel $\gamma(t-t')$ by the standard fluctuation-dissipation relation, $\langle f_i(t) f_j(t')\rangle=k_BT\gamma(t-t') \delta_{ij}$ where $i=x,y$ and $j=x,y$. Using the Laplace transformation the solution of the equations of motion can be written as

\begin{eqnarray}\label{leqs}
g_1(t)&=& x(t)-<x(t)>\nonumber\\
&=&\int_0^t H_0(t-\tau)f_x(\tau)d\tau-\Omega^2\int_0^t H_0^\prime(t-\tau)f_x(\tau)d\tau+\Omega\int_0^t H(t-\tau)f_y(\tau)d\tau , \nonumber\\
g_2(t)&=& y(t)-<y(t)>\label{leqs1}\\
&=&\int_0^t H_0(t-\tau)f_y(\tau)d\tau-\Omega^2\int_0^t H_0^\prime(t-\tau)f_y(\tau)d\tau-\Omega\int_0^t H(t-\tau)f_x(\tau)d\tau , \nonumber\\
g_3(t)&=& {u_x}(t)-<{u_x}(t)>\label{leqs2}\\
&=&\int_0^t \dot{H_0}(t-\tau)f_x(\tau)d\tau-\Omega^2\int_0^t \dot{H_0^\prime}(t-\tau)f_x(\tau)d\tau+\Omega\int_0^t \dot{H}(t-\tau)f_y(\tau)d\tau , \nonumber\\
g_4(t)&=& {u_y}(t)-<{u_y}(t)>\label{jj}\\
&=&\int_0^t \dot{H_0}(t-\tau)f_y(\tau)d\tau-\Omega^2\int_0^t \dot{H_0^\prime}(t-\tau)f_y(\tau)d\tau-\Omega\int_0^t \dot{H}(t-\tau)f_x(\tau)d\tau , 
\end{eqnarray}
\noindent
where
\begin{equation}\label{leqs4}
<x(t)>=A(t)x(0)-B(t)y(0)+C(t)u_x(0)+D(t)u_y(0)=c_1(say),
\end{equation}
\begin{equation}\label{leqs5}
<u_x(t)>=\dot{A}(t)x(0)-\dot{B}(t)y(0)+\dot{C}(t)u_x(0)+\dot{D}(t)u_y(0)=c_3(say),
\end{equation}
\begin{equation}\label{leqs6}
<y(t)>=A(t)y(0)+B(t)x(0)+C(t)u_y(0)-D(t)u_x(0)=c_2(say) ,
\end{equation}
\begin{equation}\label{leqs7}
<u_y(t)>=\dot{A}(t)y(0)+\dot{B}(t)x(0)+\dot{C}(t)u_y(0)-\dot{D}(t)u_x(0)=c_4(say) .
\end{equation}
\noindent
with $A\equiv A(t) =\chi_0(t)+\Omega^2\omega^2\chi(t)$, $B\equiv B(t) =\Omega\omega^2H^\prime(t)$,
$C\equiv C(t) =H_0(t)-\Omega^2H_0^\prime(t)$, $D\equiv D(t) =\Omega H(t)$,$\chi_0(t)=1-\omega^2\int_0^tH_0(\tau)d\tau$ and $\chi(t)=\int_0^t{H_0}^\prime(\tau)d\tau$.
The last two relations imply that $\chi_0(0)=1.0$ and $\chi(0)=0$. Then  using the definitions of $A(t), B(t), C(t)$ and $D(t)$ one may determine the values of $H_0(t)$, $H_0^\prime(t)$ , $H(t)$ and $H^\prime(t)$ and their time derivatives at $t=0$ comparing the left and the right sides of the appropriate equation  (in the set of Eqs.(\ref{leqs4}-\ref{leqs7})) for $\Omega=0$ and $\Omega\neq 0$, respectively. Thus $\dot{H_0}(0)=1$, $H_0^\prime(0)=H(0)=H^\prime(0)=H_0(0)=0, \dot{H_0^\prime}(0)=\dot{H}(0)=\dot{H^\prime}(0)=0$. 
\par
Now we have to define the functions $H_0(t)$, ${H_0}^\prime(t)$, $H(t)$ and $H^\prime(t)$ which appear in the above equations. These are the inverse Laplace transformation of  $\tilde{H_0}(s)$, $\tilde{H_0^\prime}(s)$, $\tilde{H}(s)$ and $\tilde{H^\prime}(s)$, respectively. Here we have used $\tilde{H}(s)=s\tilde{H^\prime}(s)$ and $\tilde{H_0^\prime}(s)=s^2\tilde{H_{00}}(s)$. $\tilde{H_0}(s)$, $\tilde{H^\prime}(s)$ and $\tilde{H_{00}}(s)$ are defined as $\tilde{H}_0(s)=\frac{1}{s^2+s\tilde{\gamma}(s)+\omega^2}$, $\tilde{H^\prime}(s)=\frac{1}{(s^2+s\tilde{\gamma}(s)+\omega^2)^2+(\Omega s)^2}$,
$\tilde{H_{00}}(s)=\frac{1}{(s^2+s\tilde{\gamma}(s)+\omega^2)[(s^2+s\tilde{\gamma}(s)+\omega^2)^2+(\Omega s)^2]}$. Here $\tilde{\gamma} (s)$ is the Laplace transform of $\gamma(t)$. 
\par
We now consider the fluctuations in position and velocity, respectively. All the second moments corresponding to the fluctuations  can be represent by the matrix, $\boldsymbol{A}^\prime(t)$ with $A_{ij}^\prime=\langle g_i(t)g_j(t)\rangle$. Using the matrix, $\boldsymbol{A}^\prime(t)$ and its inverse, $\boldsymbol{A^\prime}^{-1}(t)$ one may write the phase space distribution function for 
the Langevin equations (\ref{leq1}-\ref{eqq}). Since the equations of motion correspond to the Gaussian noise driven linear system then the phase space distribution function is a Gaussian one \cite{risken}. It can be written as 

\begin{equation} \label{eq76}
P(x, x(0); y, y(0); u_x, u_x(0); u_y, u_y(0);t) = (2\pi)^{-2} (\sigma(t))^{-\frac{1}{2}} \exp\left[-\frac{1}{2}{\bf{g}}\dag(t){\boldsymbol{A^\prime}^{-1}}(t){\bf{g}}(t)\right] \; \;,  
\end{equation}
\noindent
where $\sigma(t)=A_1A_2-A_3^2-A_4^2$ and $\bf{g}(t)$ is a column matrix with the elements $g_1(t),g_2(t),g_3(t)$ and $g_4(t))$, respectively. In the above equation we have used $A_{11}^\prime=A_{22}^\prime=A_1$, $A_{33}^\prime=A_{44}^\prime=A_2$, $A_{13}^\prime=A_{31}^\prime=A_{24}^\prime=A_{42}^\prime=A_3$, $A_{14}^\prime=A_{41}^\prime=A_4$, and $A_{23}^\prime=A_{32}^\prime=-A_4$. It is to be noted here that the rest of the off diagonal elements of the matrix, $\boldsymbol{A}^\prime(t)$ are zero. 

\noindent
Now following the procedure as reported 
in the recent paper \cite{das} for several linear systems, one may read the
Fokker-Planck equation  with the solution (\ref{eq76}) as
\begin{eqnarray} \label{eq84}
\frac{\partial P}{\partial t} &=& -\frac{\partial u_x P}{\partial x}-\frac{\partial u_y P}{\partial y}+H_1(t)\left[x \frac{\partial P}{\partial u_x}+ y \frac{\partial P}{\partial u_y}\right]+H_2(t)\left[\frac{\partial u_x P}{\partial u_x}+\frac{\partial u_yP}{\partial u_y}\right]\nonumber\\
&-&H_3(t)\left[\frac{\partial u_y P}{\partial u_x}-\frac{\partial u_x P}{\partial u_y}\right] 
+ H_4(t)\left[x \frac{\partial P}{\partial y}- y \frac{\partial P}{\partial x}\right]+H_5(t)\left[u_x \frac{\partial P}{\partial y}- u_y \frac{\partial P}{\partial x}\right]\nonumber\\
&-&H_6(t)\left[x \frac{\partial P}{\partial u_y}- y \frac{\partial P}{\partial u_x}\right]
+H_7(t)\left[\frac{\partial}{\partial x}\frac{\partial P}{\partial u_y}-\frac{\partial}{\partial y}\frac{\partial P}{\partial u_x}\right]\nonumber\\
&+&H_8(t)\left[\frac{\partial^2 P}{\partial x^2}+\frac{\partial^2 P}{\partial y^2}\right]
+H_9(t)\left[\frac{\partial}{\partial x}\frac{\partial P}{\partial u_x}+\frac{\partial}{\partial y}\frac{\partial P}{\partial u_y}\right] + H_{10}(t)\left[\frac{\partial^2 P}{\partial u_x^2}+\frac{\partial^2 P}{\partial u_y^2}\right] \; \;,
\end{eqnarray}
\noindent
where $H_1 (t),H_2 (t),H_3(t),H_4(t),H_5(t),H_6(t),H_7(t),H_8(t)$, $H_9(t)$ and $H_{10}(t)$ are relevant time dependent quantities to account the non-Markovian dynamics properly. The first two terms in the right hand side of the above equation are usual drift terms in the phase space description for both Markvian \cite{risken} and non-Markovian dynamics\cite{das,adel,wang}, respectively. The next term is corresponding to the harmonic force field\cite{das,adel,wang}. Then contribution from the dissipative force is considered by the fourth term\cite{das,adel,wang}. The next drift term may be identified as due to the magnetic force \citep{das,physha}. Although additional drift terms in the presence of a magnetic field do not appear for the Markovian dynamics\cite{risken,preab} but the non-Markovian dynamics may modify the probability flux.  Keeping it in mind and the cross effect of the magnetic force, one may include additional all possible drift and diffusion terms. Thus sixth to eighth and ninth to tenth are the additional drift and diffusion terms, respectively. It is to be noted here that the calculation of the second moment also implies to include ninth and tenth terms. If the proposed Fokker-Planck equation is a correct one then fifth to tenth terms should disappear in the absence of the magnetic field. We will check it after the determination of all the coefficients. Finally, eleven-th and twelveth terms are the usual diffusion terms in the phase space description\cite{das,adel,wang}.
To avoid any confusion we would mention here that the diffusion terms with other possible cross derivatives are not considered since the cross correlation of the fluctuations is zero for the respective case.
\par
Now we are in a position to determine all the coefficients. Then putting the distribution function  (\ref{eq76}) in the above equation, we have collected coefficients of $x$, $y$, $u_x$, $u_y$, $x^2$, $y^2$, $u_x^2$,  $u_y^2$, $xu_x$, $yu_y$, $xu_y$, $yu_x$ and sum of other terms which are not having independent variables. Each coefficient and the
sum must be zero to become the distribution function as a solution of the proposed Fokker-Planck equation. Using this condition we have
\begin{multline} \label{eqx}
H_1(t)(A_1c_3-A_3c_1+A_4c_2)+H_4(t)(A_4c_3-A_3c_4+A_2c_2)-H_6(t)(A_1c_4-A_3c_2-A_4c_1)\\
+\frac{H_7(t)}{\sigma(t)}\left[c_4(A_3^2-A_4^2-A_1A_2)+2A_2A_4c_1-2A_3A_4c_3\right]
+\frac{H_8(t)}{\sigma(t)}\left[2A_2(A_3c_3+A_4c_4-A_2c_1)\right]\\
+\frac{H_9(t)}{\sigma(t)}\left[c_3(A_4^2-A_3^2-A_1A_2)+2A_3A_2c_1-2A_3A_4c_4\right]
+\frac{H_{10}(t)}{\sigma(t)}\left[-2c_1(A_3^2+A_4^2)+2A_3A_1c_3+2A_4A_1c_4\right]\\
=\frac{\dot{\sigma}(t)}{\sigma(t)}\left[A_3c_3+A_4c_4-A_2c_1\right]-\left[\dot{A_3}c_3+A_3\dot{c_3}+\dot{A_4}c_4+A_4\dot{c_4}-\dot{A_2}c_1-A_2\dot{c_1}\right] ,
\end{multline}
\begin{multline} \label{eqy}
H_1(t)(A_1c_4-A_3c_2-A_4c_1)+H_4(t)(A_3c_3+A_4c_4-A_2c_1)+H_6(t)(A_1c_3-A_3c_1+A_4c_2)\\
+\frac{H_7(t)}{\sigma(t)}\left[c_3(A_4^2-A_3^2+A_1A_2)+2A_2A_4c_2-2A_3A_4c_4\right]
+\frac{H_8(t)}{\sigma(t)}\left[2A_2(A_3c_4-A_2c_2-A_4c_3)\right]\\
+\frac{H_9(t)}{\sigma(t)}\left[c_4(A_4^2-A_3^2-A_1A_2)+2A_3A_2c_2+2A_3A_4c_3\right]
+\frac{H_{10}(t)}{\sigma(t)}\left[-2c_2(A_3^2+A_4^2)-2A_1A_4c_3+2A_1A_3c_4\right]\\
=\frac{\dot{\sigma}(t)}{\sigma(t)}\left[A_3c_4-A_4c_3-A_2c_2\right]-\left[\dot{A_3}c_4+A_3\dot{c_4}-\dot{A_4}c_3-A_4\dot{c_3}-\dot{A_2}c_2-A_2\dot{c_2}\right] ,
\end{multline}
\begin{multline} \label{equx}
H_2(t)(A_1c_3-A_3c_1+A_4c_2)+H_3(t)(A_1c_4-A_3c_2-A_4c_1)+H_5(t)(A_4c_3-A_3c_4+A_2c_2)\\
+\frac{H_7(t)}{\sigma(t)}\left[c_2(A_4^2-A_3^2+A_1A_2)-2A_3A_4c_1+2A_1A_4c_3\right]
+\frac{H_8(t)}{\sigma(t)}\left[-2c_3(A_3^2+A_4^2)+2A_3A_2c_1-2A_4A_2c_2\right]\\
+\frac{H_9(t)}{\sigma(t)}\left[c_1(A_4^2-A_3^2-A_1A_2)+2A_1A_3c_3+2A_3A_4c_2\right]
+\frac{H_{10}(t)}{\sigma(t)}\left[-2A_1(A_1c_3-A_3c_2+A_4c_2)\right]\\
=\frac{\dot{\sigma}(t)}{\sigma(t)}\left[A_3c_1-A_1c_3-A_4c_2\right]-\left[\dot{A_3}c_1+A_3\dot{c_1}-\dot{A_1}c_3-A_1\dot{c_3}-\dot{A_4}c_2-A_2\dot{c_4}\right]\\
-\left[A_3c_3+A_4c_4-A_2c_1\right] ,
\end{multline}
\begin{multline} \label{equy}
H_2(t)(A_1c_4-A_3c_2-A_4c_1)-H_3(t)(A_1c_3-A_3c_1+A_4c_2)+H_5(t)(A_3c_3+A_4c_4-A_2c_1)\\
+\frac{H_7(t)}{\sigma(t)}\left[c_1(A_3^2-A_4^2+A_1A_2)+2A_1A_4c_4-2A_3A_4c_2\right]
+\frac{H_8(t)}{\sigma(t)}\left[-2c_4(A_3^2+A_4^2)+2A_4A_2c_1-2A_3A_2c_2\right]\\
+\frac{H_9(t)}{\sigma(t)}\left[c_2(A_4^2-A_3^2-A_1A_2)-2A_4A_3c_1+2A_3A_1c_4\right]
+\frac{H_{10}(t)}{\sigma(t)}\left[-2A_1(A_1c_4-A_3c_2-A_4c_1)\right]\\
=\frac{\dot{\sigma}(t)}{\sigma(t)}\left[-A_1c_4+A_3c_2+A_4c_1\right]-\left[\dot{A_3}c_2+A_3\dot{c_2}-\dot{A_1}c_4-A_1\dot{c_4}+\dot{A_4}c_1+A_4\dot{c_1}\right]\\
-\left[A_3c_4-A_4c_3-A_2c_2\right] ,
\end{multline}
\begin{multline} \label{eqxx}
H_1(t)A_3-H_6(t)A_4-\frac{H_7(t)}{\sigma(t)}A_2A_4+\frac{H_8(t)}{\sigma(t)}A_2^2-\frac{H_9(t)}{\sigma(t)}A_3A_2+\frac{H_{10}(t)}{\sigma(t)}(A_3^2+A_4^2)\\=\frac{\dot{\sigma}(t)A_2}{2\sigma(t)}-\frac{\dot{A_2}}{2} \; \;,
\end{multline}
\begin{multline} \label{equxux}
-A_3-H_2(t)A_1-H_5(t)A_4-\frac{H_7(t)}{\sigma(t)}A_1A_4+\frac{H_8(t)}{\sigma(t)}(A_3^2+A_4^2)-\frac{H_9(t)}{\sigma(t)}A_3A_2+\frac{H_{10}(t)}{\sigma(t)}A_1^2\\=\frac{\dot{\sigma}(t)A_1}{2\sigma(t)}-\frac{\dot{A_1}}{2} ,
\end{multline}
\begin{multline} \label{eqxux}
A_2-H_1(t)A_1+H_2(t)A_3+H_3A_4-H_4(t)A_4+\frac{H_7(t)}{\sigma(t)}2A_3A_4-\frac{H_8(t)}{\sigma(t)}2A_2A_3-\frac{H_9(t)}{\sigma(t)}(A_3^2-A_4^2+A_1A_2)\\-\frac{H_{10}(t)}{\sigma(t)}2A_1A_3=-\frac{\dot{\sigma}(t)A_3}{\sigma(t)}+\dot{A_3} ,   
\end{multline}
\begin{multline} \label{eqxux}
H_2(t)A_4-H_3A_3+H_4(t)A_3+H_5(t)A_2+H_6(t)A_1+\frac{H_7(t)}{\sigma(t)}(A_4^2-A_3^2+A_1A_2)-\frac{H_8(t)}{\sigma(t)}2A_2A_4+\frac{H_9(t)}{\sigma(t)}2A_3A_4\\-\frac{H_{10}(t)}{\sigma(t)}2A_1A_4=-\frac{\dot{\sigma}(t)A_4}{\sigma(t)}+\dot{A_4}  ,
\end{multline}
\noindent
and 
\begin{multline} \label{norm}
2H_2(t)\sigma(t)+2H_7(t)A_4-2H_8(t)A_2+2H_9(t)A_3-2H_{10}(t)A_1=-\dot{\sigma}(t) .
\end{multline}
\noindent
To avoid any confusion it is to be noted here that we have above nine independent relations instead of thirteen as a consequence of the following fact. The coefficients for the pair, ($x^2, y^2$) are same. Similarly, it is true for the pairs, ($u_x^2, u_y^2$), ($xu_x^2, yu_y^2$) and ($u_x^2, u_y^2$), respectively. Thus we have nine independent relations among ten unknown coefficients of the Fokker-Planck equation. Then we need an additional condition. In this context the comparison between the Fokker-Planck description in the phase space and the configuration space for the Brownian motion of the Harmonic oscillator may give an important suggestion. The diffusion term in the configuration space does not appear in the phase space description for both Markovian\cite{risken} and non-Markovian \cite{das,adel} descriptions. Considering this we have chosen $H_8=0$. This choice will be naturally justified if the distribution function (\ref{eq76}) is a solution of the Fokker-Planck equation with the remaining terms. Using $H_8=0$ in the above set of equations one may define nine coefficients as  $H_1=\frac{\left[-\ddot{A}a_x(t)-\ddot{B}a_y(t)-\ddot{C}a_{v_x}(t)-\ddot{D}a_{v_y}(t))\right]}{\Delta_m}$, $H_2=\frac{\left[\ddot{A}b_x(t)-\ddot{B}b_y(t)-\ddot{C}b_{v_x}(t)-\ddot{D}b_{v_y}(t))\right]}{\Delta_m}$ , $H_3=\frac{\left[-\ddot{A}d_x(t)+\ddot{B}d_y(t)-\ddot{C}d_{v_x}(t)+\ddot{D}d_{v_y}(t))\right]}{\Delta_m}$, $H_4=0$, $H_5=0$, $H_6=\frac{\left[-\ddot{A}c_x(t)+\ddot{B}c_y(t)+\ddot{C}c_{v_x}(t)-\ddot{D}c_{v_y}(t))\right]}{\Delta_m}$, $H_7=\left[\dot{A_4}+H_2A_4+H_3A_3-H_6A_1\right]$, $H_9=\left[\dot{A_3}-A_2-H_3A_4+H_1A_1+H_2A_3\right]$ and $H_{10}=\frac{1}{2}\left[\dot{A_2}+2H_1A_3+2H_2A_2-2H_6A_4\right]$. Here we have used $\Delta_m= (A^2+B^2)(\dot{C}^2+\dot{D}^2)+(C^2+D^2)(\dot{A}^2+\dot{B}^2)-2(AC-BD)(\dot{A}\dot{C}-\dot{B}\dot{D})-2(AD+BC)(\dot{A}\dot{D}+\dot{B}\dot{C})$, $a_x(t)=A(\dot{C}^2+\dot{D}^2)-C(\dot{A}\dot{C}-\dot{B}\dot{D})-D(\dot{A}\dot{D}+\dot{B}\dot{C})$, $c_x(t)=B(\dot{C}^2+\dot{D}^2)+D(\dot{A}\dot{C}-\dot{B}\dot{D})-C(\dot{A}\dot{D}+\dot{B}\dot{C})$, $b_x(t)=B(C\dot{D}-\dot{C}D)+C(A\dot{C}-\dot{A}C)+D(A\dot{D}-D\dot{A})$,$d_x(t)=B(C\dot{C}+D\dot{D})-C(A\dot{D}+C\dot{B})+D(A\dot{C}-D\dot{B})$, $a_y(t)=B(\dot{C}^2+\dot{D}^2)+D(\dot{A}\dot{C}-\dot{B}\dot{D})-C(\dot{A}\dot{D}+\dot{B}\dot{C})$, $c_y(t)=A(\dot{C}^2+\dot{D}^2)-C(\dot{A}\dot{C}-\dot{B}\dot{D})-D(\dot{A}\dot{D}+\dot{B}\dot{C})$, $b_y(t)=A(C\dot{D}-\dot{C}D)-C(B\dot{C}-\dot{C}B)-D(B\dot{D}-\dot{B}D)$, $d_y(t)=A(C\dot{C}+\dot{D}D)+C(B\dot{D}-\dot{A}C)-D(B\dot{C}+\dot{A}D)$, $a_{v_x}(t)=C(\dot{A}^2+\dot{B}^2)-A(\dot{A}\dot{C}-\dot{B}\dot{D})-B(\dot{A}\dot{D}+\dot{B}\dot{C})$, $c_{v_x}(t)=D(\dot{A}^2+\dot{B}^2)+B(\dot{A}\dot{C}-\dot{B}\dot{D})-A(\dot{A}\dot{D}+\dot{B}\dot{C})$, $b_{v_x}(t)=A(A\dot{C}-\dot{A}C)+B(B\dot{C}-\dot{B}C)-D(A\dot{B}-\dot{A}B)$, $d_{v_x}(t)=A(A\dot{D}+\dot{B}C)+B(B\dot{D}-\dot{A}C)-D(A\dot{A}+\dot{B}B)$, $a_{v_y}(t)=D(\dot{A}^2+\dot{B}^2)+B(\dot{A}\dot{C}-\dot{B}\dot{D})-A(\dot{A}\dot{D}+\dot{B}\dot{C})$,
$c_{v_y}(t)=C(\dot{A}^2+\dot{B}^2)-A(\dot{A}\dot{C}-\dot{B}\dot{D})-B(\dot{A}\dot{D}+\dot{B}\dot{C})$,
$b_{v_y}(t)=A(A\dot{D}-\dot{A}D)+B(B\dot{D}-\dot{B}D)+C(A\dot{B}-\dot{A}B)$ and $d_{v_y}(t)=A(A\dot{C}-\dot{B}D)+B(B\dot{C}+\dot{A}D)-C(A\dot{A}+\dot{B}B)$. Thus the determination of the coefficients automatically implies that the distribution function (\ref{eq76}) is a solution of the Fokker-Planck equation, 
\begin{eqnarray} \label{eq84}
\frac{\partial P}{\partial t} &=& -{\bf{u}}.\nabla_{\bf{x}}P+H_1(t){\bf{x}}.\nabla_{\bf{u}}P+H_2(t)\nabla_{\bf{u}}.{\bf{u}}P\nonumber\\
&+&H_3(t)\left[{\bf{u}}\times\nabla_{\bf{u}}P\right]_z\nonumber\\
&-&H_6(t)\left[{\bf{x}}\times\nabla_{\bf{u}}P\right]_z+ H_7(t)\left[\nabla_{\bf{x}}\times\nabla_{\bf{u}}P\right]_z\nonumber\\
&+&H_9(t)\nabla_{\bf{u}}.\nabla_{\bf{x}}P
+H_{10}(t){\nabla_{\bf{u}}}^2P 
\end{eqnarray}
\noindent
Using the definition of the coefficients one may check easily that the distribution function ({\ref{eq76}}) is a solution of the above Fokker-Planck equation. It constitutes the necessary and sufficient check of the present calculation.  
Now we have to compare the above equation with the Fokker-Planck equation  which is derived recently in Ref.\cite{preha} for the same equations of motion and the associated distribution function. Then one can easily find out that the FPE in \cite{preha} contains additional three
terms with the coefficients, $H_4, H_5$ and $H_8$, respectively. The remaining terms exactly correspond with each other. At this circumstance our check of the coefficients, $H_4$, $H_5$ and $H_8$ in the respective Fokker-Planck equation in Ref.\cite{preha} suggests that $H_4=H_5=H_8=0$.  Thus  taking care of all the comments (including typo and the rearrangement of the Fokker-Planck  equation) in Sec.V (in Ref.\cite{preha}) which is devoted for Ref.\cite{das} we conclude that both the methods give the same result. Another point is to be noted here. From the independent relations among the time dependent coefficients the present method automatically requires that one of the coefficients in Eq.(\ref{eq84}) must be zero. Then we have chosen that the coefficient in the diffusion term (which appears in the Fokker-Planck equation in the configuration space) may be zero. Because it is well known that this term does not appear usually\cite{das,adel,wang,physha,preab} in the probabilistic description in velocity space or phase space for the Gaussian noise driven dynamical systems. With this choice the present method predicts automatically other coefficients exactly as the distribution function satisfies the above equation. To derive the same equation, the method \cite{preha} with the characteristic function does not
need such kind of any choice which may offer a shortcut way for the same destination (as shown in the present case). In other words, all the terms in Eq.(\ref{eq84}) and other case appear automatically in \cite{preha}.     
But the above discussion does not mean that the present method always need to include a choice as like as the present case. For examples one may go through the Ref.\cite{das}. Finally, to avoid any confusion we would mention here that if any choice appears in the method as like in the present case that may not be an arbitrary one as mentioned above. 

 Before leaving this issue we would mention that the above equation reduces to the standard results at the appropriate limits such as at $\Omega=0$. At this limit 
$A(t)=\chi_0(t)$, $B(t)=0$, $C(t)=H_0(t)$, $D(t)=0$ and $A_4=0$. Then
$H_3=H_6=H_7=0$ and the remaining  coefficients are given by $H_1(t) = \tilde{\omega}^2(t)$, $H_2(t) = \tilde{\beta}(t)$, $H_9(t) = \frac{k_BT}{\omega^2}\left[\tilde{\omega}^2(t)-\omega^2\right]$
and $H_{10}(t) = k_BT\tilde{\beta}(t)$. Here we have used $\tilde{\beta}(t) = -\frac{d\ln\Delta^\prime(t)}{dt}$, $\tilde{\omega}^2(t) = \frac{\dot{A}\ddot{C}-\dot{C}\ddot{A}}{\Delta^\prime(t)}$ and $\Delta^\prime(t) = \left[A\dot{C}-\dot{A}C\right]$. Thus at $\Omega=0$, the Fokker-Planck equation (\ref{eq84})reduces to the standard result\cite{adel,wang}. 

For further check, we consider the condition, $\omega=0$. Then
$A\equiv A(t) =1$, 
$B\equiv B(t) =0$,  
$C\equiv C(t) =H_0(t)-\Omega^2H_0^\prime(t)$, 
$D\equiv D(t) =\Omega H(t)$,
$\chi_0(t)=1$ and 
$\chi(t)=\int_0^tH_0^\prime(\tau)d\tau$. Here
$H_0(t)$, $H_0^\prime(t)$, $H(t)$ and $H^\prime(t)$ are the inverse Laplace transformation of  $\tilde{H_0}(s)=\frac{1}{s^2+s\tilde{\gamma}(s)}$, $\tilde{H_0^\prime}(s)=s^2(\frac{1}{(s^2+s\tilde{\gamma}(s))[(s^2+s\tilde{\gamma}(s))^2+(\Omega s)^2]})$, $\tilde{H}(s)=s\tilde{H^\prime}(s)$ and $\tilde{H^\prime}(s)=\frac{1}{(s^2+s\tilde{\gamma}(s))^2+(\Omega s)^2}$, respectively.
Then one can easily show that $H_1=H_4=H_5=H_6=0$. Thus in the absence of the harmonic force field the Fokker-Planck Eq.(\ref{eq84}) reduces to 
\begin{eqnarray} \label{eq84b}
\frac{\partial P}{\partial t} &=& -{\bf{u}}.\nabla_{\bf{x}}P+H_2(t)\nabla_{\bf{u}}.{\bf{u}}P\nonumber\\
&+&H_3(t)\left[{\bf{u}}\times\nabla_{\bf{u}}P\right]_z\nonumber\\
&+& H_7(t)\left[\nabla_{\bf{x}}\times\nabla_{\bf{u}}P\right]_z\nonumber\\
&+&H_9(t)\nabla_{\bf{u}}.\nabla_{\bf{x}}P
+H_{10}(t){\nabla_{\bf{u}}}^2P 
\end{eqnarray}

\noindent
with
$H_2=-\frac{\ddot{C}\dot{C}+\ddot{D}\dot{D}}{\dot{C}^2+\dot{D}^2}$,
$H_3=\frac{\ddot{D}\dot{C}-\ddot{C}\dot{D}}{\dot{C}^2+\dot{D}^2}$,
$H_7=\left[\dot{A_4}+H_2A_4+H_3A_3\right]$,
$H_9=\left[\dot{A_3}-A_2-H_3A_4+H_2A_3\right]$, 
$H_{10}=\frac{1}{2}\left[\dot{A_2}+2H_2A_2\right]$ and
$\Delta_m=(\dot{C}^2+\dot{D}^2)$.

It is to be noted here that  
the above equation corresponds to the limiting case (absence of time dependent force fields) of the Fokker-Planck equation which was derived in Ref.\cite{physha} using the characteristic function.  Thus the accuracy of the present method is well justified with  the check of the calculation for appropriate limiting conditions. Very recently, using it the Fokker-Planck equation has been derived in Ref.\cite{das1} for the non-Markovian dynamics in the presence of magnetic field and time dependent conservative force. This equation reduces to all the standard results at appropriate limits. Thus the present method may be applicable for any kind of linear Langevin equation of motion which describes additive colored noise driven non Markovian dynamics with or without frictional memory kernel.

\par
In conclusion, the present calculation suggests that the drift terms, $H_4(t)\left[x \frac{\partial P}{\partial y}- y \frac{\partial P}{\partial x}\right]$, $H_5(t)\left[u_x \frac{\partial P}{\partial y}- u_y \frac{\partial P}{\partial x}\right]$
and the diffusion term, $H_8(t)\left[\frac{\partial^2 P}{\partial x^2}+\frac{\partial^2 P}{\partial y^2}\right]$ are not relevant quantities in the Fokker-Planck description of the Brownian motion of a Harmonic oscillator in the presence of a magnetic field and the non-Markovian thermal bath. At the same, it contradicts the claim made in Ref.\citep{preha} in the context of comment on Ref.\cite{das}. 
The authors in \citep{preha} claimed that
their method is accurate compared to the calculation by Das 
{\it et al.}\cite{das}. In other words, the present calculation justifies that both the methods give the same result.   

\end{document}